\documentclass[preprint,12pt]{elsarticle}

\usepackage{graphicx}

\journal{Physica A}
\begin{document}
\begin{frontmatter}
\title{Planar unclustered scale-free graphs as models for technological and biological networks.
}

\author{Alicia Miralles, Francesc Comellas\corref{corresponding}}
\ead{ [almirall,comellas]@ma4.upc.edu}
\cortext[corresponding]{Corresponding author.  PMT Edifici C3, c/ Esteve Terradas 5,
08860 Castelldefels, Barcelona, Catalonia, Spain,Tel. +34 934 134 109 Fax +34 934 137 007}
\address{
Departament de Matem\`atica Aplicada IV, Universitat Polit\`ecnica de Catalunya\\
08860 Castelldefels, Catalonia, Spain}
\author{Lichao Chen, Zhongzhi Zhang}
\ead{zhangzz@fudan.edu.cn}
\address{
School of Computer Science and Shanghai Key Lab of Intelligent Information Processing,\\
Fudan University, Shanghai 200433, China}%
%

\begin{abstract}
Many real life networks present an average path length logarithmic with 
the number of nodes and a degree distribution which follows a power law. 
Often these networks have also a modular and self-similar structure and,  
in some cases - usually associated with topological restrictions- their clustering is low
and they are almost planar.
In this paper we introduce a family of  graphs which share all these properties and are 
defined by  two parameters. As their construction is deterministic, we obtain 
exact analytic expressions 
for relevant properties of the graphs including the
degree distribution, degree correlation, diameter, and average distance, as a
function of the two defining parameters. Thus, the graphs are useful to model some complex networks,
in particular several families of  technological and biological networks, and
in the design of new practical communication algorithms in relation to their dynamical processes.
They can also help understanding the underlying mechanisms that have produced their particular structure.
\end{abstract}
\end{frontmatter}

\section{Introduction}
Ten years have past since the publication of the groundbreaking papers by 
Watts and Strogatz~\cite{WaSt98} on small-world networks and 
Baraba«si and Albert~\cite{BaAl99} on scale-free networks.
Their works led researchers to the design of new network models to describe 
complex systems in nature and society like the Internet,  protein-protein interactions, 
transportation systems or social and economic networks. 
Their models try to match observational studies which have identified  at least three 
important common characteristics  for real-life networks:
They exhibit a small average distance and diameter (compared to a random network 
with the same number of nodes and links);
the number of links attached to the nodes obeys a power-law distribution 
(the networks are  scale-free); and  recently it has been discovered that, often,
real networks are self-similar~\cite{SoHaMa05} and show a degree hierarchy related to the modularity of the system, see~\cite{RaSoMoOlBa02,SaGuMoNu07,ClMoNe08} .

Many of the proposed models are stochastic as this is the case for the now classical 
preferential attachment method~\cite{BaAl99}. 
Thus, the use of mean field techniques is required to estimate the main 
parameters of a network~\cite{BaAlJe99}. 
However, a deterministic approach has proven useful to complement
and enhance the probabilistic and simulation techniques.
Deterministic models have a clear advantage, as  they
allow an analytical exact determination of relevant network parameters, 
which then can be compared with experimental data coming 
from real and simulated networks

Among the different methods known to generate deterministic models
those based on recursive or iterative methods are of particular interest.
In these methods, new nodes are added and connected to a given
substructure of the network at each generation step.
This is the case for  pseudo-fractal networks~\cite{DoGoMe02}
where, at each step,  new vertices are added simultaneously, one to each already existing link.
This  construction can be generalized if  complete subgraphs of a given size (cliques)
are considered instead of links (which are of course 2-cliques), see~\cite{CoFeRa04}.
Similar rules give the Apollonian networks~\cite{AnHeAnSi05,DoMa05,ZhCoFeRo06}.
On the other hand, there also exist techniques that produce  networks by duplication of a given 
substructure, see~\cite{ChLuDeGa03,SoPaSmKe01}.

A generalization of  these two methods introduces at each iteration  a 
substructure which is added to the network, according to a deterministic rule. 
Substructures that have been used are triangles~\cite{ZhZhFaGuZh07}, 
cycles~\cite{CoZhCh09} and paths~\cite{CoMi09}.

In this paper we go one step further  by considering the simultaneous introduction of 
$d$  substructures in parallel -in our case, paths- which are attached to the same basic unit (a link)
 generalizing the model given in~\cite{CoMi09}, which added a single path to each link.
The resulting graphs are essentially different from those in~\cite{CoMi09}. 
In particular they are scale-free (with a power-law exponent which depends on $d$) 
while in~\cite{CoMi09} the degree distribution is exponential.
The model is a family of planar, modular, hierarchical and self-similar
networks, with small-world scale-free characteristics and
with clustering coefficient zero, and all these parameter are
determined by $d$ as well as by the iteration step $t$.
We note that some important real life networks, for example  those associated to
electronic circuits, Internet and some biological systems~\cite{FeJaSo01,Ne03},
have these characteristics as they are modular,
almost planar and with a reduced clustering coefficient and have
small-world and scale-free properties.
Thus, these networks are modeled by our  construction which can be considered 
as a new tool in the study of their associated complex systems.
In particular, the model could be used to find also practical algorithms in relation to dynamical processes 
(synchronization, cover time, etc.) for these technological and biological networks 
and can help  understanding the underlying mechanisms that have produced their particular structure.

In the next section we introduce the family of graphs object of  study and in Section~\ref{sec:topo} 
we calculate analytically some relevant properties
for the graphs, namely,  the degree distribution, degree correlations,  the diameter and
the average distance.  The last section provides some conclusions.

\section{Generation of the graphs $M_d(t)$}

In this section we introduce a family of modular, self-similar and
planar graphs which have the  small-world pro\-per\-ty and are scale-free.
The family depends on an adjustable parameter  $d$  and the iteration
number $t$.
We provide an iterative algorithm,  and also a recursive method, for
its construction.
The construction methods allow  a direct determination of the number of vertices (nodes) and edges (links) of the graph.

\medskip{\em Iterative construction.--}
We give here an iterative formal definition of the proposed family
of graphs, $M_d(t)$, characterized by  $t\ge 0$, the number of
iterations and a parameter $d$ associated with the self-repeating modular structure.

First, we call {\em generating edges}  the only edge of $M_d(0)$ and
all edges of $M_d(t)$ whose endvertices have  been introduced at
different  iteration steps $t$. All other edges of $M_d(t)$ will be
known as  {\em passive edges}.
A generating edge becomes passive after its use in the construction.

The graph $M_d(t)$ is constructed as follows:

For $t=0$, $M_d(0)$ has two vertices and a generating  edge
connecting them.

For  $t\geq 1$, $M_d(t)$ is  obtained from $M_d(t-1)$ by adding,  to
every generating edge in $M_d(t-1)$,  $d$ parallel paths of length three (each path has four vertices and three edges)
 by identifying  the two final vertices of each path  with the endvertices of the generating edge.

\begin{figure}[htbp]
\begin{center}\includegraphics[scale=0.35]{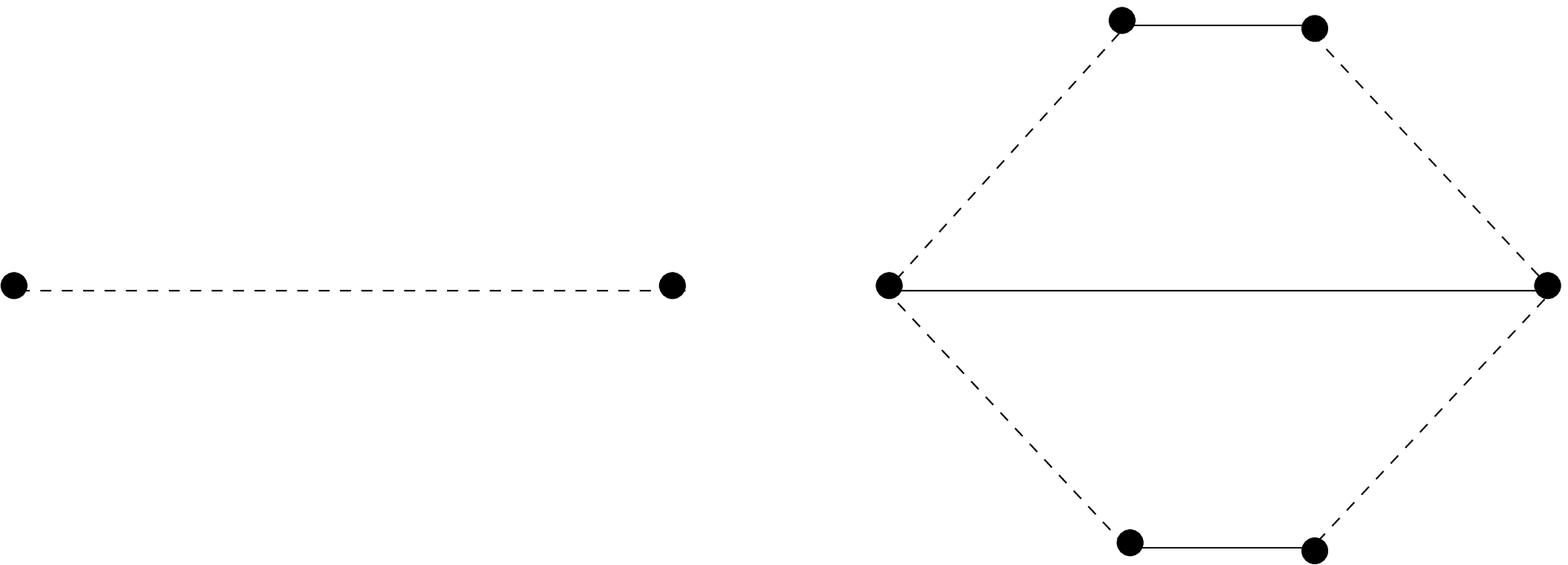}\end{center}
\vskip 0.5cm
\begin{center}
\includegraphics[scale=0.6]{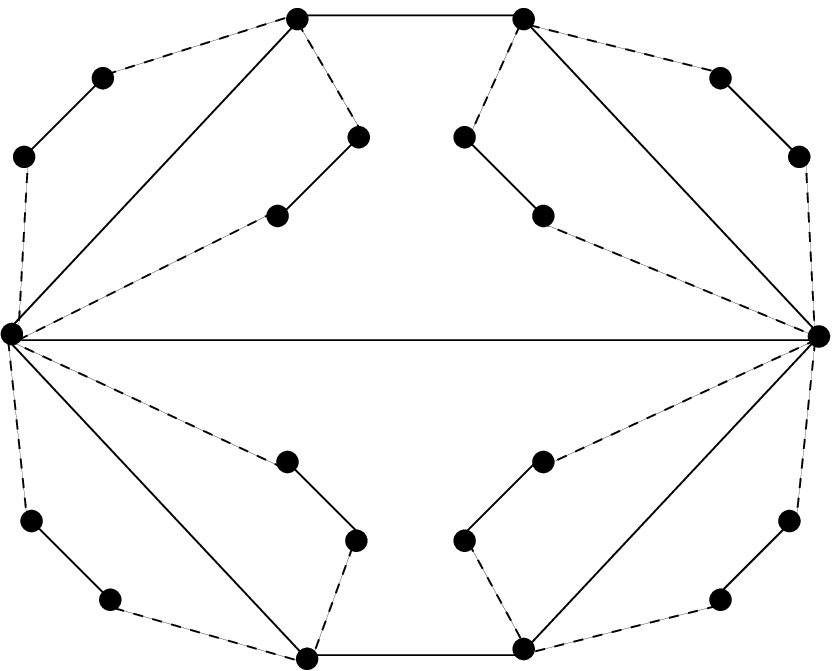}
\end{center}
\begin{center}
\vskip 0.5cm
\includegraphics[scale=0.6]{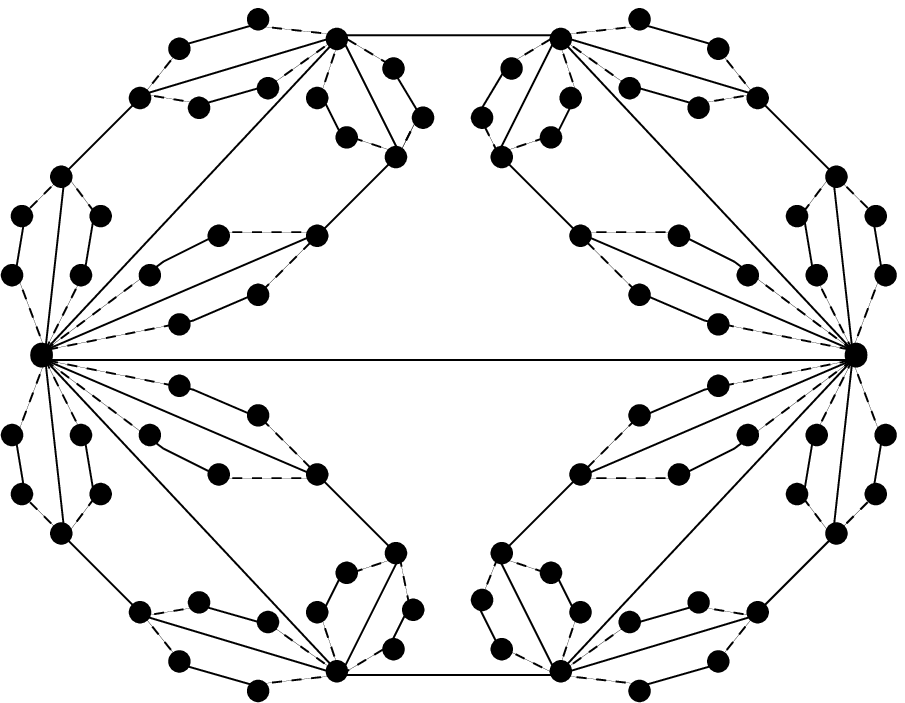}
\end{center}
\caption{Graphs $M_d(t)$ produced at  iterations $t=0, 1, 2$ and
$3$ for $d=2$.} \label{fig:recmod}
\end{figure}

The process is repeated until the desired number of vertices is reached, see Fig.~\ref{fig:recmod}.
We note that the number of vertices can be also adjusted with the parameter $d$ (number of parallel paths that are 
attached to each generating edge).

\medskip{\em Recursive modular construction.--}
The graph $M_d (t)$ can  also be defined as follows:

\noindent (a) For $t=0$, $M_d (0)$ has two vertices and a generating edge connecting them.

\noindent (b) For  $t=1$, $M_d (1)$ is obtained  from $M_d (0)$ by adding to its only edge 
 $d$ parallel paths  of length three
 by identifying  the two final vertices of each path with
 the endvertices of the initial edge. 

\noindent (c)  For  $t\geq 2$,  $M_d (t)$  is made from $2d$ copies of  $M_d (t-1)$,
by identifying, vertex to vertex,  the initial edge of each $M_d (t-1)$  with the generating edges
of  $M_d (1)$,  see  Fig.~\ref{fig:recmod}.


\medskip{\em Number of vertices and edges of $M_d(t)$.--}
We use the following notation: $\tilde{V}(t)$, $\tilde{E}(t)$ and
$\tilde{E_g}(t)$ denote, respectively,  the set of  vertices,
edges and generating edges introduced at step $t$, while $V(t)$
and $E(t)$ denote the set of vertices and edges of the graph
$M_d(t)$.

Notice that, at each iteration, a generating edge is replaced by
$2d$ new generating edges and $d$ passive edges. Therefore:
$|\tilde{E_g}(t+1)|=2d\cdot|\tilde{E_g}(t)|$, and
$|\tilde{E_g}(t)|=(2d)^t$. As each generating edge introduces at
the next iteration $2d$ new vertices and $3d$ new edges we have
$|\tilde{V}(t+1)|=2d\cdot|\tilde{E_g}(t)|=(2d)^{t+1}$ and
$|\tilde{E}(t+1)|=3d\cdot|\tilde{E_g}(t)|=3d\cdot (2d)^{t}$. As
$|\tilde{V}(0)|=2$ and $|\tilde{E_g}(0)|=1$, the number of vertices and edges of
$M(t)$, $t\ge 0$, is:
\begin{eqnarray}\label{OrdSiz}
|V(t)|&=&\sum_{i=0}^t |\tilde{V}(i)|=\frac{(2d)^{t+1}+2d-2}{2d-1},\nonumber \\
|E(t)|&=&\sum_{i=0}^t |\tilde{E}(i)|= \frac{3d(2d)^t-d-1}{2d-1}.
\end{eqnarray}
%

\medskip{\em Planarity.--} A graph is planar if it
can be drawn on the plane with no edges crossing.
By construction of $M_d(t)$, the
introduction at each iteration of $d$ parallel paths connected to each generating
edge, which afterwards becomes passive, adds $2d$ new vertices to
the graph and they can be drawn without crossing edges.
Planarity could also be proven from  Kuratowski's theorem or 
from the  known planarity test which states that a graph is planar if
it has no cycles of length 3 and $|E|\leq 2|V|-4, |V|>3$, see ~\cite{Di05}.

\section{Topological properties of $M_d(t)$}\label{sec:topo}
Thanks to the deterministic nature of the graphs $M_d(t)$, we can give exact
values for  the relevant topological pro\-per\-ties of this graph family, namely,
the degree distribution, degree correlations, the diameter and the average distance.

\medskip{\em Degree distribution.--}
Initially, at $t=0$,  the graph has two vertices of degree one.
When a new vertex $i$ is added to the graph at iteration $t_i$,
this vertex has degree $2$ and it is connected to only one
generating edge. We use the following notation: $k_g(i,t)$,
$k_p(i,t)$ and $k(i,t)$ are, res\-pectively, the number of
generating edges, passive edges and total edges
connected to  vertex $i$, at step $t\ge t_i$. 
Therefore $k(i,t)=k_g(i,t)+k_p(i,t)$  is  the degree of vertex $i$ at this step.

From the construction process we can write,
\begin{equation}\label{recurrenciagrau}
\left \{\begin{array}{ll} k_p(i,t+1)=k_p(i,t)+k_g(i,t) \\
k_g(i,t+1)=d  k_g(i,t)\end{array}\right .
\end{equation}
with the initial conditions,
\begin{eqnarray}
k_g(i,t_i)&=&1, \quad  t_i  \geq 0\quad \textnormal{and} 
\quad \nonumber \\ 
k_p(i,t_i)&=&\left \{\begin{array}{ll} 0 \quad\textnormal{if}\quad
t_i=0 \quad  \\ 1 \quad\textnormal{otherwise}\end{array}\right .
\end{eqnarray}
and for $d>1$ we have,
\begin{eqnarray}
k_g(i,t)&=&d^{t-t_i}, \quad  t_i  \geq 0 \quad \textnormal{and}
\quad \nonumber \\ 
k_p(i,t)&=&\left \{\begin{array}{ll}
1+\frac{d^{t}-d}{d-1}
\quad\textnormal{if}\quad t_i=0 \quad  \\
2+\frac{d^{t-t_i}-d}{d-1}
\quad\textnormal{otherwise.}\end{array}\right .
\end{eqnarray}

All the vertices that have been introduced at step $t_i$ 
have the same degree at step $t$:
\begin{enumerate}
\item The two vertices introduced at step $t_i=0$  have degree,
\begin{eqnarray} \label{degreeti=0}
k(i,t)&=&k_g(i,t)+k_p(i,t)=\nonumber\\
&=&d^{t}+1+\frac{d^{t}-d}{d-1}=\frac{d^{t+1}-1}{d-1}.
\end{eqnarray}
\item The $|\tilde{V}(t_i)|=(2d)^{t_i}$ vertices introduced at step $t_i>0$  have degree,
\begin{eqnarray}\label{degreeti>0}
k(i,t)&=&k_g(i,t)+k_p(i,t)=\nonumber\\
&=&d^{t-t_i}+2+\frac{d^{t-t_i}-d}{d-1}=1+\frac{dd^{t-t_i}-1}{d-1}.
\end{eqnarray}
\end{enumerate}

Therefore the  degree spectrum of the graph is discrete and
to relate the exponent of this discrete degree 
distribution to the power law  exponent of a continuous degree distribution 
for random scale free networks, 
we use the technique described by  Newman in~\cite{Ne03} to find
the cumulative degree distribution $P_{\rm cum}(k)$. If we denote
by $V(t,k)$ the set of vertices that have degree $k$ at step $t$,
\begin{eqnarray*}
P_{\rm cum}(k)&=&\frac{\sum_{k'\ge k}|V(t,k')|}{|V(t)|}
    =\frac{2+\sum_{t'_i=1}^{t_i}(2d)^{t'_i}}{\frac{(2d)^{t+1}+2d-2}{2d-1}}=\\
    &=& \frac{(2d)^{t_i+1}+2d-2}{(2d)^{t+1}+2d-2}=\\
    &=&\frac{(2d)^{t-\frac{\ln (k+\frac{2-k}{d}-1)}{\ln
    (d)}+1}+2d-2}{(2d)^{t+1}+2d-2}.
\end{eqnarray*}

Fot  $t$ large, we obtain,

\begin{eqnarray*}
P_{\rm cum}(k)&\approx &(2d)^{-\frac{\ln (k+\frac{2-k}{d}-1)}{\ln (d)}}=(k+\frac{2-k}{d}-1)^{-\frac{\ln(2d)}{\ln (d)}}\\
&=& k^{-\frac{\ln(2d)}{\ln
(d)}}(1-\frac{1}{d}+\frac{2-d}{kd})^{-\frac{\ln(2d)}{\ln (d)}}.
\end{eqnarray*}

\begin{figure}[htbp]
\begin{center}
\includegraphics[scale=0.7]{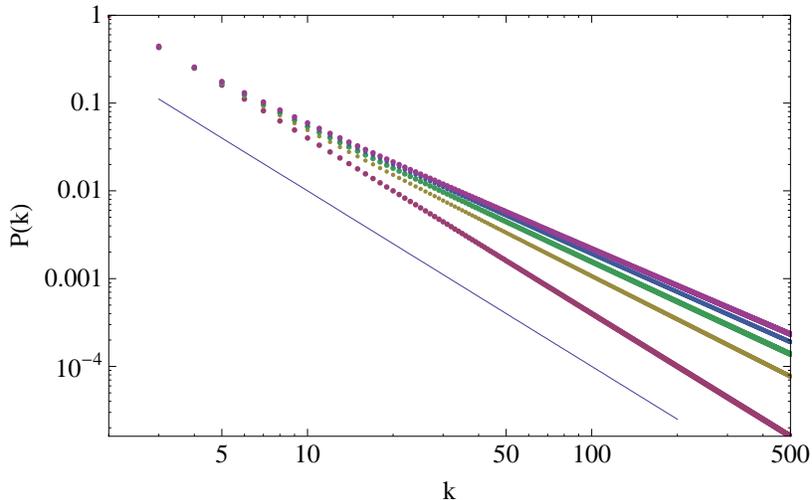}
\caption{Log-log representation of the cumulative degree distribution for  $M_d(t)$,  $d=2, 3, 4, 5$. 
The reference line has slope -2.} \label{fig:degree}
\end{center}
\end{figure}

For $k>>1$ this expression gives
\begin{equation}
P_{\rm cum}(k)\approx k^{-\frac{\ln(2d)}{\ln
(d)}}(1-\frac{1}{d})^{-\frac{\ln(2d)}{\ln (d)}}
\end{equation}
Thus, the degree distribution follows a power-law
$$P_{\rm cum}(k)\sim k^{-\gamma}, \;\;  \mbox{with   }  \gamma =\frac{\ln(2d)}{\ln (d)},$$ 
and therefore the degree distribution is scale-free, see Fig.~\ref{fig:degree}.

Research on networks associated to
electronic circuits show that many of them are almost planar,
modular and have a small clustering coefficient and in most cases
their degree distributions follow a power-law~\cite{FeJaSo01,Ne03} with
exponent values in the same range than those of $M_d(t)$.

\medskip{\em Correlation coefficient.--}
We have obtained  the Pearson correlation coefficient~\cite{Ne02}, $r(d,t)$,  for
the degrees of the endvertices of the edges of $M_d(t)$.
In Appendix A we present the details of the calculation that leads to
the  its  exact  analytical expression as shown in Eq.~\ref{pearson}.
We particularize this general analytical result for different instances of the graphs,  obtaining  numerical 
values of the correlation as shown in Table~\ref{tab:correlacio} .
\begin{table}[htbp]
\begin{center}
\begin{tabular}{c|c|c|c|c}
\hline
 & $t=1$  & $t=2$  & $t=3$  & $t=10$ \\
\hline\hline

& & & \\
$d=2$ & $-0.1667$ & $-0.0886$ & $-0.0460$ & $-0.0003$ \\
\hline
$d=10$ & $-0.4091$ & $-0.2338$ & $-0.1174$ & $-0.0009$\\
\hline
$d=100$ & $-0.4901$ & $-0.2057$ & $-0.0934$ & $-0.0007$ \\
\hline\hline
\end{tabular}
 \end{center}
\caption{Correlation coefficient at steps $t=1,2,3,10$ for
se\-ve\-ral values of $d$.} \label{tab:correlacio}
\end{table}

From the analytical results and the  numerical values of the correlation coefficient  we see that this family of graphs has the
degrees of the endvertices negatively correlated (large degree vertices tend to be connected 
with low degree vertices) and the graphs are disassortative, as it occurs with many technological and 
biological networks~\cite{Ne03}.

For $d>>1$,  we obtain $\displaystyle{r(d,t)\approx 
\frac{1}{1-3\cdot 2^{t-1}}},$
which for $t$ large gives $ r(d,t)\sim 0$.

\medskip{\em Diameter.--} At each iteration step we
introduce, for every generating edge, $2d$ new vertices. These vertices
are among them at distance at most 3.
As each vertex  joins the graph of the former step through
one new edge, the diameter will increase by exactly 2 units.
Therefore $D(t)= D(t-1)+2$, $t\geq 2$.  As  $D(1)=3$, we have that
the diameter of  $M_d(t)$ is $D(t)=3+2\cdot (t-1)$,  $t\geq 1$. Therefore,
from Eq.~\ref{OrdSiz}, and as for $t$ large, $t\sim \ln |V(t)|$ we
have in this limit that $D(t)\sim \ln |V(t)|$.

\medskip{\em Average distance.--} 
The average distance of $M_d(t)$ is
defined as:
\begin{equation}\label{apl01}
\bar{D}(t)  =  \frac{1}{{\mbox{\scriptsize $ |V(t)| (|V(t)|-1)/2$}}}
\sum_{i,j \in V(t)} d_{i,j} \,,
\end{equation}
where $d_{i,j} $ is the distance between vertices $i$ and $j$. 

In Appendix B we use the modular recursive construction of $M_d(t)$ to
calculate the exact value of  $\bar{D}(t) $ which results:
\begin{eqnarray}
\bar{D}(t)&=&(-1 + 4 d - 5 d^2 + 2 d^3 + 2^{1 + t} d^{1 + t} -  7\cdot 2^{2 t} d^{2 + 2 t} + \nonumber\\
    &+& 3\cdot 2^{1+2 t} d^{3+2 t}-2^{1+t} d^{1+t} t + 3\cdot 2^{1+t} d^{2+t}t  -\nonumber\\
  &-&2^{2+t} d^{3+t} t-2^{1+2 t} d^{2+2 t} t+2^{2+2 t} d^{3+2 t} t
  )\nonumber\\
   &/& ((-1 + 2 d) (-1 + d + 2^t d^{1 + t})(-1 + 2^{1 + t}d^{1 + t})).\nonumber\\
 \end{eqnarray}

Notice that for a large iteration step, $t \rightarrow \infty$,
$\bar{D}(t) \simeq t \sim \ln |V(t)|$, which  shows a
logarithmic scaling of  the average distance with the number of vertices of the
graph. As we have a similar behavior for  the diameter, the
graph is small-world.
\section{$M_d(t)$ as a model for some technical and biological networks}

The graphs introduced here have parameters which are similar to those of some real life networks.
A good  example is  the largest  benchmark considered in \cite{FeJaSo01} --a network
with 24097 nodes, 53248 edges, average degree 4.34 and average
distance 11.05--  has a degree distribution which follows a power-law
with exponent 3.0, and it has a small clustering coefficient $C=0.01$ and 
other network properties  are also in the same range than those of  the graph $M_6(4)$, see~\cite{Ne03}.  
Table \ref{tab:comparison} compares some network parameters from instances of our model with data coming from real networks published elsewhere.
Although there are many similarities between the two sets, the aim of this model is not to match perfectly all the network parameters for some real life complex systems, but to provide an analytical framework where to perform precise tests of new algorithms (routing, synchronization, etc.) and check properties that otherwise would require less general and precise techniques like simulation of stochastic methods.


\begin{table*}[htbp]
\footnotesize
\begin{center}
\begin{tabular}{l|r|r|r|r|r|r|r|r}
\hline
Network  & Vertices & Edges & $\gamma$ & Avg.   & Clust. &  Avg.   &r & Ref(s). \\
                &                 &              &                     &  dist. &              &  degree &  &  \\

\hline 
$M_2(5)$  &  1366  & 2047 &       3 & 6.850 &  0 & 2.997  & -0.001& \\
{\em Java Dev. Fram.} &  1376  & 2174 & 2.5 & 6.39 &  0.06 & 3.160 & -0.002 & ~\cite{VaFeSo02,Ne03}\\
\hline
$M_6(2)$  &  158  & 235 & 2.39 & 3.290 &  0 & 2.975  & -0.233 & \\
{\em Silwood Pk food web} &  154  & 366 & 1 & 3.4 &  0.15 & 4.75  & -0.31 & ~\cite{SoVa04}\\
\hline 
$M_6(3)$  &  1886  & 2827 & 2.39 & 4.474 &  0 & 2.998  & -0.130 & \\
{\em protein inter. S.C.}  &  2115  & 2240 & 2.4 & 6.80 &  0.071 & 2.089 & -0.156 & ~\cite{JeMaBaOl01,Ne03}\\
\hline
$M_6(4)$  &  22622  & 33931 & 2.39 & 5.557 &  0 & 3.000  & -0.007 & \\
{\em electronic circuits}  &  24097  & 53248 & 2.39 & 11.05 &  0.01 & 4.34  & -0.130 & ~\cite{FeJaSo01,Ne03}\\
\hline
$M_8(3)$  &  4370  & 6553 & 2.33 & 4.482 &  0 & 2.999  & -0.123 & \\
{\em power grid}  &  4941  & 6594 &   & 19.99 &  0.1 & 2.669 & -0.003 & ~\cite{WaSt98,Ne03}\\
\hline
\end{tabular}
 \end{center}
\caption{Some instances of $M_d(t)$ and possible real network counterparts.} \label{tab:comparison}
\end{table*}

\section{Conclusion}
The graphs  $M_d(t)$ introduced and studied here are
planar, modular, have a disassortative degree hierarchy  and are  small-world and scale-free.
Another relevant characteristic of the  graphs is their zero clustering. 
A combination of a low clustering coefficient, modularity, and small-world scale-free
properties  can be  found in some real networks,  in particular in  technological 
and biological networks~\cite{Ne03,FeJaSo01}, and  most of them are also
disassortative.

Finally, we should emphasize that the planar property and  the deterministic character of the
family, in contrast with more usual probabilistic approaches, should facilitate the exact determination 
of other network parameters and the development of  new network algorithms that then might be
extended to real-life complex systems.

\subsection*{Acknowledgments}
F. Comellas and A. Miralles are supported by the 
Ministerio de Ciencia e Innovaci\'on, Spain, and the European Regional Development Fund under project MTM2008-06620-C03-01 and partially supported by the Catalan Research Council under grant 2009SGR1387.
L. Chen and Z. Zhang are supported by the National Natural Science Foundation of China under  Grant No. 60704044.
\vskip 1cm

\noindent{\bf APPENDICES}


\appendix
\section{Correlation coefficient calculation.}
The Pearson correlation coefficient, $r(d,t)$,  for
the degrees of the endvertices of the edges of $M_d(t)$ is:
\begin{equation}\label{rNe}
r(d,t)=\frac{|E(t)|\sum_i j_i k_i -[\sum_i \frac{1}{2} (j_i+ k_i)]^2
}{|E(t)|\sum_i \frac{1}{2} (j_i^2+ k_i^2) -[\sum_i
\frac{1}{2}(j_i+ k_i)]^2 }
\end{equation}
where $j_i$, $k_i$ are the degrees of the endvertices of the $i$th
edge, with $i=1,\cdots ,|E(t)|$, see~\cite{Ne02} .

To calculate  the correlation coefficient we need to know the degree distribution of the
endvertices of  the edges in $\tilde E(t_i)$ at a given step
$t_i$. We denote by $\langle j,k\rangle$ an edge connecting
vertices of degrees $j$ and $k$.

The detail  of this distribution is given as follows:

The edges introduced at step $t_i$ are:

\begin{enumerate}
\item Edges $\langle 2,2\rangle$, connecting two vertices introduced at step $t_i>0$. There are ${(2d)^{t_i}}/{2}$ edges
(a half of the vertices introduced at step $t_i$). Notice that
there is one edge $\langle 1,1\rangle$ introduced at $t_i=0$.

\item Edges $\langle 2,k(i',t_i)\rangle$ connecting vertices of degree two, introduced at step $t_i$,
with all the vertices $i'$  introduced at step $t_{i'}$ with $0\le t_{i'}\le
t_i-1$. For each vertex  $i'$ there are
$k_g(i',t_i)$ edges: 

From the two vertices introduced at $t_{i'}=0$,   see
(\ref{degreeti=0}),    there are $2 d^{t_{i'}}$ edges $\langle
2,\frac{d^{t_i+1}-1}{d-1}\rangle$.

From the $(2d)^{t_{i'}}$ vertices introduced at $t_{i'}>0$,    see
(\ref{degreeti>0}),  there are $(2d)^{t_{i'}} d^{t_i-t_{i'}}$
edges $\langle 2,1+\frac{d\,d^{t_i-t_{i'}}-1}{d-1}\rangle$.

\end{enumerate}

Table \ref{tab:edgdist} here displays a summary of the  results.
\begin{table*}[htbp]
\begin{center}
\begin{tabular}{c|c|c|c}
\hline
Step $t_i$  & Edges at step $t_i$  &Number  & Edges at step $t>t_i$  \\
\hline\hline
$\begin{array}{c} \\ t_i=0 \\  \\ \end{array}$ & $\langle 1,1\rangle$ & $1$ & $\langle \frac{d^{t+1}-1}{d-1},\frac{d^{t+1}-1}{d-1}\rangle$ \\
\hline
$\begin{array}{c} \\ 1\le t_i \le t \\ \\ \end{array}$ &
$\begin{array}{c} \\ \langle 2,2\rangle \\ \\ \langle
2,\frac{d^{t_i+1}-1}{d-1}\rangle \\ \\ \end{array}$ &
$\begin{array}{c}  \frac{(2d)^{t_i}}{2} \\ \\ 2d^{t_i}\\
\end{array}$  & $\begin{array}{c} \langle
1+\frac{dd^{t-t_i}-1}{d-1},1+\frac{dd^{t-t_i}-1}{d-1}\rangle \\ \\
\langle 1+\frac{dd^{t-t_i}-1}{d-1},\frac{d^{t+1}-1}{d-1}\rangle
 \\ \end{array}$ \\
\hline
$\begin{array}{c}\\ 2\le t_i\le t \\ \\1\le t_{i'}\le t_i-1\\  \\\end{array}$ & $ \langle 2,1+\frac{dd^{t_i-t_{i'}}-1}{d-1}\rangle $ & $ (2d)^{t_{i'}}\cdot d^{t_i-t_{i'}}$ & $\langle 1+\frac{dd^{t-t_i}-1}{d-1},1+\frac{dd^{t-t_{i'}}-1}{d-1}\rangle $ \\
\hline\hline
\end{tabular}
 \end{center}
\caption{Number of edges in $M_d(t)$  according to the degrees of
their endvertices.} \label{tab:edgdist}
\end{table*}

Using these  results, we can find the following sums:

\begin{small}
\begin{eqnarray*}
\sum_i j_i k_i
&=&(4+16\,d-51\,d^2+41\,d^3-8\,d^4-\nonumber\\
&-&3\,d^5+d^6+d^{t+1}(40+8t-80\cdot2^t)+\nonumber\\
&+&d^{t+2}(-184-40t+282\cdot 2^t)+\nonumber\\
&+&d^{t+3}(306+74t-373\cdot 2^t)+\nonumber\\
&+&d^{t+4}(-236-64t+227\cdot 2^t)+\nonumber\\
&+&d^{t+5}(86+26t-63\cdot 2^t)+\nonumber\\
&+&d^{t+6}(-12-4t+7\cdot 2^t)+\nonumber\\
&+&d^{2t+2}(10+4t)+d^{2t+3}(-43-18t)+\nonumber\\
&+&d^{2t+4}(62+28t)+d^{2t+5}(-35-18t)+\nonumber\\
&+&d^{2t+6}(6+4t))/((d-1)^3\nonumber\\
& &(2d^2-5d+2)(d-2)),\nonumber\\
\null & & \null \nonumber\\
\sum_i (j_i+ k_i)&=& \frac{-2}{(2d-1)(d-2)(d-1)^2}(-2-3d+\nonumber\\
&+&10d^2-6d^3+d^4+d^{t+1}(-4+16\cdot 2^t) +\nonumber\\
&+& d^{t+2}(14-37\cdot 2^t)+d^{t+3}(14+26\cdot 2^t)+\nonumber\\
&+& d^{t+4}(4-5\cdot 2^t)-d^{2t+2} +3d^{2t+3}-\nonumber\\
&-&2d^{2t+4} )\nonumber ,\\
\null & & \null \nonumber\\
\sum_i (j_i^2+ k_i^2)&=& -2( -8-32d+186d^2-282d^3+\nonumber\\
&+&145d^4+49d^5-96d^6+48d^7-11d^{8} +\nonumber\\
&+& d^{9} +d^{t+1}(-72+160\cdot 2^t)+d^{t+2}(384-\nonumber\\
&-&728\cdot 2^t)+d^{t+3}(-750+1252\cdot 2^t)+\nonumber\\
&+&d^{t+4}(606-882\cdot 2^t)+ d^{t+5}(-33-39\cdot 2^t)+\nonumber\\
&+&d^{t+6}(-285+456\cdot 2^t)+d^{t+7}(201-286\cdot 2^t)+\nonumber\\
&+& d^{t+8}(-57+74\cdot 2^t)+d^{t+9}(6-7\cdot 2^t)+\nonumber\\
&+& 4d^{3t+2} -16d^{3t+4} +17d^{3t+5}+5d^{3t+6}-\nonumber\\
&-&19d^{3t+7}+11d^{3t+8}-2d^{3t+9})/((d^2-2d+1)\nonumber\\
& &(2d-1)(d-2)(d^3-2d^2-2d+4)(d-1)^2).\\
\end{eqnarray*}
\end{small}

Replacing these sums into equation~(\ref{rNe}) we obtain,  for any $d$,  the exact  analytical expression for the Pearson correlation coefficient of $M_d(t)$ which is displayed as Eq.~\ref{pearson}.  For $d=2$ this equation becomes Eq.~\ref{pearson2}:

%
\begin{figure*}[htb]
\line(1,0){400}

\begin{equation}\label{pearson}
r(d,t)=\frac{(d(3\cdot 2^t d^t-1)-1)\cdot (B(d,t)+C(d,t))-(D(d,t)+E(d,t))^2}{(d(3\cdot 2^t d^t-1)-1)\cdot F(d,t)-(D(d,t)+E(d,t))^2}, d>2
\end{equation}

\mbox{Where:}

\footnotesize
\begin{eqnarray*}
B(d,t)&=& (d^3+2\cdot d^2-6d-1)+d^{t+1}2(2d-1)((5-3d)+\nonumber\\
          &+& t(1-d))+d^{t+2}(-28\cdot 2^t+(-5+2t)d^t)+   d^{t+3}(7\cdot 2^t+(6+4t)d^t),\nonumber\\
C(d,t)&=&(d^{t+3}(31\cdot 2^t+(-11+6t)d^t)+d^{t+2}(-90\cdot 2^t+\nonumber\\
              &+&(10-12t)d^t)+d^{t+1}80\cdot 2^t)/((d-2)^2),\nonumber\\
D(d,t)&=&d^2-3d-1+d^t(6\cdot 2^t-6d^t)+d^{t+1}(11\cdot 2^t-3d^t-2)+d^{t+2}(-5\cdot 2^t-2d^t+4),\nonumber\\
E(d,t)&=&-12d^t (d^t-2^t)/(d-2),\nonumber\\
F(d,t)&=&( d^{5+t}(-6+7\cdot 2^t+2d^{2t})+d^{4+t}(21-32\cdot 2^t+d^{2t})+d^{3+t}(3+3\cdot 2^t-d^{2t})+\nonumber\\
         &+&d^{2+t}(-42+62\cdot 2^t)+ d^{1+t}(18-40\cdot 2^t)-d^5+5d^4-5d^3-11d^2+14d+2 )/(d^2-2).\nonumber\\
\end{eqnarray*}\nonumber

\mbox{For $d=2$:}


\begin{equation}\label{pearson2}
r(2,t)=\frac{4^t t^2-2^{t+1} t+3\cdot 2^{2 t+1} t-2^{3 t+2} t+13\cdot 4^t-3\cdot 2^{t+1}+4^{2 t+1}-3\cdot 2^{3 t+2}+1}{2^{4 t+1}
   t^2+2^{2 t+1} t-2^{3 t+3} t+2^{4 t+3} t-2^t+5\cdot 4^t-2^{3 t+4}+3\cdot 2^{4 t+3}-3\cdot 2^{5 t+2}}
\end{equation}\label{eq:corr2}

\line(1,0){400}

\end{figure*}
\normalsize


\section{Analytical determination of the average distance.}
The average distance of $M_d(t)$ is
defined as:
\begin{equation}\label{apl01}
\bar{D}(t)  =  \frac{1}{{\mbox{\scriptsize $ |V(t)| (|V(t)|-1)/2$}}}
\sum_{i,j \in V(t)} d_{i,j} \,,
\end{equation}
where $d_{i,j} $ is the distance between vertices $i$ and $j$. In
what follows, $S(t)$ will denote the sum  $ \sum_{i,j \in V(t)}
d_{i,j}$.


\begin{figure}[htb]
\begin{center}
\includegraphics[scale=1.2]{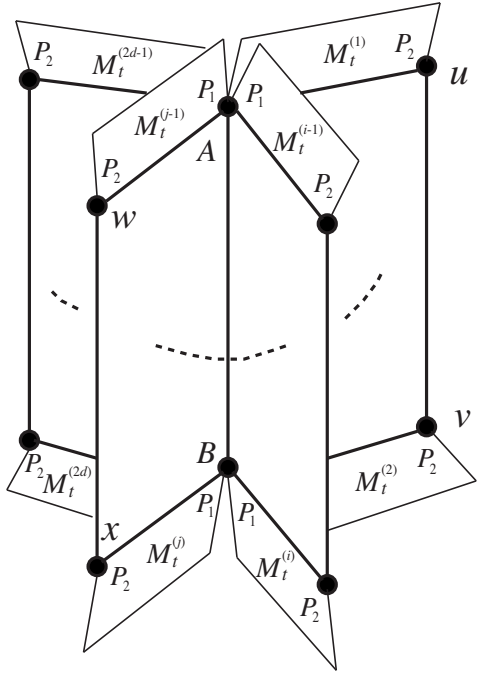}
\end{center}
\caption{$M_d(t+1)$ is obtained from the juxtaposition of $2d$
copies of $M_d(t)$.}\label{fig:llibre}
\end{figure}

The modular recursive construction of $M_d(t)$ allows us to
calculate the exact value of  $\bar{D}(t) $.
 At step $t$, $M_d(t+1)$ is obtained from the juxtaposition of $2d$ copies of
$M_d(t)$,  which we label $M_{d,t}^{(\eta)}$, $\eta=1,2,\cdots,
2d$, see Figures~\ref{fig:recmod} and~\ref{fig:llibre}. 
Whenever possible, we drop the
subscript $d$ and represent $M_{d,t}^{(\eta)}$ as $M_{t}^{(\eta)}$
to keep the notation uncluttered.  
The copies are connected one to
another at  $2d+2$ vertices which we call {\em connecting vertices}.
Two of them are the initial two vertices of the graph, 
which will be denoted in this section as $A$ and $B$.

In Fig.~\ref{fig:llibre} we display $A$, $B$ and four more of these vertices, denoted 
as $u$, $v$, $w$ and $x$. 
Note that in this figure, and for the sake of clarity, each copy
of $M_d(t)$ has been represented as a rectangle, and only its connecting 
vertices have been drawn.

Thus, the sum of distances $S_{t+1}$ satisfies the following recursion:
\begin{equation}\label{apl03}
  S_{t+1} = 2d\, S_t + \Delta_t.
\end{equation}
where $\Delta_t$ is the sum over all shortest path length whose
endpoints are not in the same $M_t^{(\eta)}$ branch.

To compute $\Delta_t$, we classify the vertices of
$M_d(t+1)$ into two categories: the two vertices with the largest degree
(i.e., $A$ and $B$ in Fig.~\ref{fig:llibre}) are called {\em hubs},
while all other vertex are called non-hub vertex. Thus
$\Delta_t$ can be obtained by
adding the following path lengths that are not included in the
distance between vertex pairs of $M_{t}^{(\eta)}$: length of the shortest
paths between non-hub vertices, length of the shortest paths between
a hub  and non-hub vertices, and length of the shortest paths
between hubs  (for example, $d_{uv}$, $d_{uB}$, and $d_{ux}$).

Let us denote $\Delta_t^{\alpha,\beta}$ as the sum of all shortest paths
between non-hub vertices, whose endpoints are in $M_t^{(\alpha)}$ and
$M_t^{(\beta)}$, respectively. Thus,
$\Delta_t^{\alpha,\beta}$ rules out the paths with endpoints at the
connecting vertices belonging to $M_t^{(\alpha)}$ or $M_t^{(\beta)}$.
For example, each path contributing to $\Delta_t^{1,2}$ does not
end at vertex $u$, $v$, $A$ or $B$, and each path contributing to
$\Delta_t^{1,4}$ does not end at vertex $u$, $A$, $B$ or $x$.
According to its value, $\Delta_t^{\alpha,\beta}$ can be split
into three classes, where the three representatives are
$\Delta_t^{1,2}$, $\Delta_t^{1,3}$, and $\Delta_t^{1,4}$, and the
cardinality of the three classes are $d$, $d(d-1)$, and $d(d-1)$,
respectively. Analogously, the length of the shortest paths between a
hub  and all non-hub vertices can be classified into two classes,
while the shortest paths between  hubs  can be partitioned
into three classes with  path lengths equal to 1, 2, or 3.

Let $\Omega_t^{\alpha}$ be the set of non-hub vertices in
$M_t^{(\alpha)}$, then the total sum $\Delta_t$ is given by
\begin{eqnarray}\label{cross01}
\Delta_t &=&d\Delta_t^{1,2}+d(d-1)
\left(\Delta_t^{1,3}+\Delta_t^{1,4}\right)+2d(d+1)\nonumber\\
& &\sum_{j \in \Omega_t^{2}}d_{Aj}+2d(d-1)\sum_{j \in
\Omega_t^{4}}d_{uj}+\nonumber
\\&+&
(d+1)d_{uv}+d(d+1)\,d_{uw}+d(d-1)\,d_{ux},
\end{eqnarray}
where $d_{uv}=1$, $d_{uw}=2$, and $d_{ux}=3$ are easily seen.

Having $\Delta_t$ in terms of the quantities of $\Delta_t^{1,2}$,
$\Delta_t^{1,3}$, $\Delta_t^{1,4}$, $\sum_{j \in
\Omega_t^{2}}d_{Aj}$ $\sum_{j \in \Omega_t^{2}}d_{uj}$, and
$\sum_{j \in \Omega_t^{2}}d_{uj}$, the next step is to explicitly
determine these quantities. To this end, we classify non-hub vertices
in  $M_d(t+1)$ into two different parts according to their
shortest path lengths to either of the two hubs (i.e. $A$ and
$B$). Notice that the vertices $A$ and $B$  themselves are not
partitioned into either of the two parts represented as $P_{1}$
and $P_{2}$, respectively. The classification of vertices is shown in
Fig.~\ref{fig:llibre}). For any non-hub vertex $\varphi$, we
denote the shortest path length from $\varphi$ to $A$, $B$ as
$a$, and $b$, respectively. By construction, $a$ and $b$ can
differ  at most by $1$ since vertices $A$ and $B$  are adjacent.
Then the classification function $class(\varphi)$ of vertex
$\varphi$ is defined to be
\begin{equation}\label{classification}
class(\varphi)=\left\{
\begin{array}{lc}
{\displaystyle{P_{1}}}
& \quad \hbox{for}\ a<b,\\
{\displaystyle{P_{2}}}& \quad \hbox{for}\ a>b.\\
\end{array} \right.
\end{equation}

It should be mentioned that the definition of the vertex classification is
recursive. For instance, class $P_{1}$ and $P_{2}$ in $M_t^{(1)}$
belong to class $P_{1}$ in $M_d(t+1)$, class $P_{1}$ and $P_{2}$ in
$M_t^{(2)}$ belong to class $P_{2}$ in $M_d(t+1)$, and so on. Since
the two hubs $A$ and $B$ are symmetrical, in the graph we
have the following equivalent relations from the viewpoint of class
cardinality: classes $P_{1}$ and $P_{2}$ are equivalent one to another. 
We denote the number of vertices in network $M_d(t)$ that belong
to class $P_{1}$ as $N_{t,P_{1}}$,  and the number of vertices in class
$P_{2}$ as $N_{t,P_{2}}$. By symmetry, we have
$N_{t,P_{1}}=N_{t,P_{2}}$, which will be abbreviated as $N_t$
hereafter. It is easy to see that
\begin{equation}\label{Np01}
N_t=\frac{|V(t)|}{2}-1=\frac{d(2d)^t-d}{2d-1}.
\end{equation}

For a vertex $\varphi$ in  $M_d(t+1)$, we are also interested
in the smallest value of the shortest path length from $\varphi$
to either of the two hubs $A$ and $B$. We denote
the shortest distance as
this value by  $f_\varphi$,  and it can be defined as
\begin{equation}\label{fv01}
f_\varphi = min(a,b).
\end{equation}

Let $\delta_{t,P_{1}}$ ($\delta_{t,P_{2}}$) denote the sum of $f_\varphi$ for
all vertices belonging to class $P_{1}$ ($P_{2}$) in $M_d(t)$.
Again by symmetry, we have $\delta_{t,P_{1}}=\delta_{t,P_{2}}$ that will be
written as $\delta_t$ for short. Taking into account the recursive method
of constructing $M_d(t)$, we notice that the vertex classification
follows also a recursion. Therefore  we can write the following
recursive formula for $\delta_{t+1}$:
\begin{equation}\label{dp01}
\delta_{t+1} = 2d\,\delta_{t}+d\,N_{t} + d.
\end{equation}
Substituting equation~(\ref{Np01}) into equation~(\ref{dp01}), and
considering the initial condition $\delta_{0}=0$, equation~(\ref{dp01})
is solved inductively
\begin{equation}\label{dp02}
\delta_{t}=\frac{2d-2 d^2-(2d)^{1+t}+d(2d)^{1+t} - dt(2d)^t+
dt(2d)^{1+t}}{2 (2 d-1)^2}.
\end{equation}

We now return to compute equation~(\ref{cross01}).
For convenience, we use $\Gamma_t^{\eta,i}$ to denote the set of
non-hub vertices belonging to class $P_i$ in $M_{t}^{(\eta)}$. Then
$\Delta_t^{1,2}$ can be written as
\begin{eqnarray}\label{cross02}
  \Delta_t^{1,2}& =& \sum_{\stackrel{r \in \Gamma_t^{1,1}\bigcup \Gamma_t^{1,2}}{s \in
      \Gamma_t^{2,1} \bigcup \Gamma_t^{2,2}}} d_{rs}\nonumber \\
      & =&\sum_{\stackrel{r \in \Gamma_t^{1,1}}{s \in
      \Gamma_t^{2,1} }} (d_{rA}+d_{AB}+d_{Bs})+\sum_{\stackrel{r \in \Gamma_t^{1,1}}{s \in
      \Gamma_t^{2,2} }} (d_{rA}+d_{Av}+d_{vs})\nonumber \\&\quad&+\sum_{\stackrel{r \in \Gamma_t^{1,2}}{s \in
      \Gamma_t^{2,1} }} (d_{ru}+d_{uB}+d_{Bs})+\sum_{\stackrel{r \in \Gamma_t^{1,2}}{s \in
      \Gamma_t^{2,2} }} (d_{ru}+d_{uv}+d_{vs})\nonumber \\&
      =&8N_t \delta_t+6(N_t)^2.
\end{eqnarray}
Analogously, we find
\begin{equation}\label{cross03}
  \Delta_t^{1,3}=8N_t\delta_t+4(N_t)^2
\end{equation}
and
\begin{equation}\label{cross04}
  \Delta_t^{1,4}=8N_t\delta_t+8(N_t)^2.
\end{equation}

Next we will determine other quantities in equation~(\ref{cross01}),
with $\sum_{j \in \Omega_t^{2}}d_{Aj}$ given by
\begin{eqnarray}\label{cross05}
\sum_{j \in \Omega_t^{2}}d_{Aj}& =&\sum_{j \in
      \Gamma_t^{2,1}} (d_{AB}+d_{Bj})+\sum_{j \in
      \Gamma_t^{2,2}} (d_{Av}+d_{vj})\nonumber \\
      & =&2\,\delta_t+3\,N_t.
\end{eqnarray}
Analogously, we can obtain
\begin{equation}\label{cross06}
 \sum_{j \in \Omega_t^{4}}d_{uj}=2\,\delta_t+5\,N_t.
\end{equation}

Substituting equations~(\ref{cross02}), (\ref{cross03}),
(\ref{cross04}), (\ref{cross05}) and (\ref{cross06}) into equation
(\ref{cross01}), we have the final expression for cross distances
$\Delta_t$,
\begin{eqnarray}\label{cross07}
 \Delta_t&=&1+5 d^2+4 d (4 d-1) N_t+6 d (2 d-1)(N_t)^2+\nonumber \\
 &+& 8 d \delta_t [d+(2 d-1) N_t]=\nonumber \\
      & =&\frac{1}{(1-2 k)^2} (1-4 d+5 d^2-2 d^3+(1-d)(2d)^{2+t}+\nonumber \\
      &+& (5+2 t)4^{1+t} d^{4+2 t}-(7+2t)d^2(2d)^{1+2 t} ).\nonumber \\
\end{eqnarray}

Inserting equation~(\ref{cross07}) into equation~(\ref{apl03}) and
using the initial condition $S_{0} =1$, equation~(\ref{apl03}) is
solved inductively,
\begin{eqnarray}\label{cross08}
S_t &=& \frac{1}{(-1+2 d)^3}
    ( -1+4 d-5 d^2+2 d^3+2^{1+t} d^{1+t}-\nonumber\\
  &-& 7\cdot 2^{2 t} d^{2+2 t}+3\cdot 2^{1+2 t} d^{3+2 t}-2^{1+t} d^{1+t} t+\nonumber\\
  &+& 3\cdot 2^{1+t} d^{2+t}t-2^{2+t} d^{3+t} t-2^{1+2 t} d^{2+2 t} t+\nonumber\\
  &+& 2^{2+2 t} d^{3+2 t} t
  ).
 \end{eqnarray}
Substituting equation~(\ref{cross08}) into  equation~(\ref{apl01})
yields the exact analytic expression for the average distance of $M_d(t)$ as
\begin{eqnarray}
\bar{D}(t)&=&(-1 + 4 d - 5 d^2 + 2 d^3 + 2^{1 + t} d^{1 + t} -  7\cdot 2^{2 t} d^{2 + 2 t} + \nonumber\\
    &+& 3\cdot 2^{1+2 t} d^{3+2 t}-2^{1+t} d^{1+t} t + 3\cdot 2^{1+t} d^{2+t}t  -\nonumber\\
  &-&2^{2+t} d^{3+t} t-2^{1+2 t} d^{2+2 t} t+2^{2+2 t} d^{3+2 t} t
  )\nonumber\\
   &/& ((-1 + 2 d) (-1 + d + 2^t d^{1 + t})(-1 + 2^{1 + t}d^{1 + t})).\nonumber\\
 \end{eqnarray}



\end{document}